\documentclass{article}
\usepackage[backref=page]{hyperref}
\usepackage{graphicx, float, amsmath,cleveref, array, caption}
\usepackage{algorithm}
\usepackage{algpseudocode}
\usepackage{wrapfig}
\usepackage{multirow}
\usepackage[numbers,compress]{natbib}
\usepackage{parskip}
\usepackage{xcolor}
\usepackage{booktabs}
\usepackage{appendix}
\usepackage{authblk}
\usepackage{optidef}

\title{Accelerated Patient-Specific Hemodynamic Simulations with Hybrid Physics-Based Neural Surrogates}
\author[1]{Natalia L. Rubio}
\author[1, 2]{Eric F. Darve}
\author[1, 2, 3,4* ]{Alison L. Marsden}
\affil[1]{Stanford University - Department of Mechanical Engineering}
\affil[2]{Stanford University - Institute for Computational and Mathematical Engineering}
\affil[3]{Stanford University - Department of Bioengineering}
\affil[4]{Stanford University - Department of Pediatrics - Cardiology}
\affil[*]{Corresponding author: amarsden@stanford.edu}
\date{}
\usepackage[total={6in,9in}]{geometry}
\Crefname{equation}{Eq.}{Eqs.}

\begin{document}
\maketitle

\section*{Abstract}
\textbf{Introduction:}

Physics-based 0D reduced-order models provide computationally lightweight predictions of cardiovascular flows, resolving bulk hemodynamics that would take days to solve using traditional 3D finite-element techniques in fractions of a second.  However, the accuracy of 0D models is limited as a result of the dramatic simplifications made in their derivations.  In this work, we use 0D parameters learned from high-fidelity 3D data to improve 0D model accuracy without sacrificing its low computational cost or interpretability.

\textbf{Methods:}

We use the resistor-quadratic resistor-inductor (RRI) and resistor-inductor (RI) model to predict pressure drops over 0D vessels and bifurcations, where the resistances and inductance (0D parameters) are predicted from the bifurcation or vessel geometry using neural networks.  The neural networks are trained on pairs of geometric input features and optimal 0D parameters extracted from high-fidelity 3D simulations.  We introduce geometric preprocessing steps to split 3+ outlet junctions into a series of bifurcations and extend the outlet boundaries into the outlet vessels by a pseudo entrance-length.  We also consider an alternative loss for neural network training that counteracts imbalances in the training data set and experiment with the exclusion of the quadratic term in the RRI model.  We validate the hybrid physics-based, data-driven framework and test the effect of these design choices for simulating hemodynamics in three types of patient-specific vasculatures - aortic, aortofemoral, and pulmonary anatomies.

\textbf{Results:}

Use of learned 0D parameters reduces error by at least $50\%$ compared to baseline 0D parameters for each anatomy-specific cohort.  The improvements are especially marked for the more complex pulmonary anatomies, where 0D models with learned parameters reduced error from 30.4\% to 7.9\%.  The quadratic resistor in the RRI model was shown to make the framework less robust in a data-scarce landscape, and did not significantly improve accuracy. The entrance length adjustment and alternative loss significantly improved results on the pulmonary anatomies with extensive branching, while having smaller effects on the simpler aortic and aortofemoral geometries.

\textbf{Conclusions:}

We present a framework for including parameters learned from high-fidelity 3D simulations in physics-based 0D models, thereby improving their accuracy.  The resulting hybrid model presents a means of real-time, interpretable, and accurate cardiovascular flow modeling, enabling digital twins that support clinical decision-making as well as cardiovascular science and engineering research.

\section{Introduction}

In recent decades, numerical models of cardiovascular hemodynamics have become a valuable tool for the study and management of cardiovascular disease.  Acting as digital twins that can be modified and interrogated without causing harm to human subjects, these simulations provide clinicians with insights for treatment decisions and surgical planning\cite{Bluestein2017UtilizingClinic, Morris201ComputationalMedicine, Schwarz2023BeyondDisease}, guide engineers in the design of medical devices\cite{Farah2025ComputationalAneurysms, Hsu2025ConstrainedGrafts, Blum2025OversizedGrafts, Hu2025MultiphysicsConduit, Fraser2011TheDevices, Gundert2012OptimizationDynamics, hossain12mathematical, Kung2013PredictiveCases}, and reveal biomechanical phenomena that further our understanding of cardiovascular physiology\cite{Brown2024ComputationalDiseases, Rolf-Pissarczyk2025MechanismsModels, Baumler2024LongitudinalDynamics,Baumler2025AssessmentModels, Dong2021ComputationalDefects}.

Traditionally, three-dimensional (3D) finite element models are used to simulate cardiovascular flows.  These models discretize the 3D vascular domain into small cells over which the Navier-Stokes equations are solved.  While this simulation methodology is highly accurate, resolving detailed velocity and pressure fields across the anatomy, it also carries a high computational cost, generally requiring days or hours of runtime on high-performance computing clusters \cite{Pfaller2022AutomatedFlow}.  The resources and time required to run these simulations limit their utility in scenarios requiring a real-time response or many-query applications such as uncertainty quantification or optimization.

Physics-based reduced-order models (ROMs) are derived by applying simplifying assumptions to the Navier-Stokes equations to arrive at a simpler system of equations \cite{Ventre2020Reduced-OrderPathologies, Formaggia2010CardiovascularSystem}.  In this work, we focus on one such ROM type, zero-dimensional (0D) models, which discretize the vasculature into a network of vessel and junction elements.  Each element is assigned a set of ``lumped parameters'' that govern its hemodynamics.  0D models can be solved in fractions of a second on ubiquitous computing equipment, such as a personal laptop or smartphone \cite{Pfaller2022AutomatedFlow, Menon2025SvZeroDSolver:Simulations}.  As such, 0D models are an important tool in cardiovascular flow modeling, despite their reduced accuracy.

In addition to acting as real-time stand-alone models \cite{Segers2003SystemicModel, Ghitti2022NonlinearVessels, Broome2013Closed-loopSystem, Pham2022SvMorph:Anatomies, Sankaran2020PhysicsSimulations, Hu2023AArteries, Hashemi2022RealReserve}, 0D models are frequently used in conjunction with 3D models.  Commonly, 0D models are used as boundary conditions.  That is, an anatomy of interest is modeled in 3D to provide a detailed, high-fidelity flow field, while the dynamics of the surrounding vasculature are captured by a 0D model, reducing the computational cost of the problem and capturing relevant physiology, possibly outside of the image acquisition region \cite{Kheyfets2013ConsiderationsHypertension, Johnson2011ApplicationPredictions, Clipp2009ImpedanceRespiration, Kheyfets2015Patient-specificCirculation, Ebrahimi2022SimulatingModel, Cai2024ACapacity, Brown2023AMechanics}.  0D model solutions are also used to initialize 3D simulations, providing a good initial guess that reduces the computation needed for the 3D model to reach convergence \cite{Nair2023Non-invasiveMeasurements, Pfaller2021OnSimulations}.  

Solution strategies for many-query problems also make use of 0D models.  For inverse parameter estimation problems (e.g., tuning boundary conditions to match clinical targets), 0D models may be used in a preliminary optimization to find an initial guess for the parameter value, which is then refined using 3D simulation \cite{Richter2025BayesianModels, Brown2023AMechanics, Nair2023Non-invasiveMeasurements, Nair2023HemodynamicsSimulation, Li2023AFlow, Ramazanli2025ModelingApproaches}. In uncertainty quantification, 0D solutions are used in multi-fidelity methods to reduce the number of 3D solutions needed to reach statistical convergence \cite{Fleeter2020MultilevelHemodynamics, Choi2025OnConditions, Tran2017AutomatedSimulations, Seo2020Multi-FidelityUncertainty, Seo2020TheWalls, Menon2024PersonalizedQuantification, Schiavazzi2018MULTIFIDELITYHEMODYNAMICS, Schaefer2024GlobalHaemodynamics, Zanoni2024ImprovedTechniques}.

In 0D models, flow through vessels is treated as fully developed, idealized Poiseuille pipe flow, and any junction dynamics are neglected.  This is a significant departure from the true, nonlinear nature of cardiovascular flows and introduces significant error, particularly for vasculatures featuring extensive branching (e.g., pulmonary arteries) or geometric complexity (e.g., stenosis or curvature) \cite{Pfaller2022AutomatedFlow}.  This is the primary drawback of 0D models, limiting their utility, particularly for high-stakes clinical applications.  

Data-driven ROMs provide an alternative to physics-based ROMs, training machine learning models to fit data rather than deriving simplified models from the Navier-Stokes equations \cite{Fresca2021Real-TimeModels, Pegolotti2021ModelVessels, Tenderini2025ModelInteraction, Ye2024Data-drivenSnapshots, Balzotti2022AGraft, Liang2020AAorta, Gharleghi2022TransientNetworks, Arzani2021Data-drivenOpportunities}.  In some cases, data-driven models are tailored to capture some of the physics of the problem, for example, graph neural nets that preserve the topology of the discretized vasculature or physics-informed neural networks that use governing physical laws in the loss used to train the machine learning model \cite{Buoso2019Reduced-orderDisease, Pegolotti2024LearningNetworks,Lannelongue2026PhysicsHemodynamics, Raissi2019Physics-informedEquations, Pan2024DomainModelling, Fu2023Physics-dataRegimes, Velikorodny2025DeepVessels, Maidu2025Super-resolutionNetworks,Arzani2021UncoveringNetworks}.  These models demonstrate the capacity to achieve higher accuracy than physics-based ROMs for a similarly low computational cost, but require large training datasets and often struggle to generalize to new geometries.  Furthermore, such ``black box'' techniques lack the interpretability of physics-based ROMs, a valuable asset in model interrogation and clinical adoption.

This work strives to combine physics-based and data-driven ROM techniques to produce a hybrid ROM that reaps the benefits of both approaches.  We consider a 0D model form, but use lumped parameters learned from high-fidelity 3D data to model vessels and junctions rather than parameters derived from Poiseuille flow assumptions.  In this way, we inject insights learned from 3D modeling (considered the ground-truth in this work) into the 0D model, improving its accuracy while retaining its interpretability.  The built-in physics knowledge and modularity of the 0D framework, in turn, provide generalizability and reduce training data demands.

In previous work, we proposed a hybrid model to better describe the hemodynamics of simple bifurcations in 0D junction elements and integrated this junction model in idealized synthetic vascular trees, yielding lower 0D model error \cite{Rubio2025HybridDifferences, Rubio2026Data-drivenModels, Sexton2025RapidBiomanufacturing}.  Here, we extend our methodology to successfully apply it to far more complex human vasculatures.  We further extend the junction modeling framework to vessels, producing further improvements in accuracy.

We first present the standard 0D model and introduce the proposed modification - use of neural-network predicted parameters to model hemodynamics in vessels and junctions.  These neural networks take the vessel or junction geometry as input and output the 0D lumped parameters.  We discuss the extraction of neural network training data from 3D geometries and the corresponding high-fidelity simulation results.  We then detail two pre-processing steps to modify the standard 0D vessel-junction discretization of a vasculature into a discretization compatible with our hybrid approach - splitting of 3+ outlet junctions into bifurcations and adjustment of junction-vessel boundaries to include an entrance length, such that flows are fully developed when crossing the boundary between junction and vessel.

We compare the accuracy of our hybrid 0D model with that of the standard 0D model on three different cohorts of patient-specific anatomies: aortic, aortofemoral, and pulmonary.  We use five-fold cross-validation to evaluate the performance of our hybrid 0D model on geometries unseen during training.  We observe significantly improved accuracy across all cohorts, with more marked improvement for the more complex anatomies.  Ultimately, this work presents an interpretable physics-based 0D ROM framework augmented with insights learned from 3D simulation to yield improved accuracy.

\section{Models and Methods}
\subsection{0D Physics-Based ROM}
0D models represent 3D vasculatures as networks of vessel and junction elements, resolving bulk flow rate $Q$ and average pressure $P$ at the nodes of the vasculature (locations where vessel and junction elements meet)\cite{Formaggia2010CardiovascularSystem, Pfaller2022AutomatedFlow, Mirramezani2019ReducedArteries, Kim2010DevelopingFlow, Menon2025SvZeroDSolver:Simulations}.  Between the inlet and outlet of a vessel, mass conservation,
 \begin{equation}
     Q_\text{V,in}-Q_\text{V,out} = 0, \quad Q_\text{V,in}=Q_\text{V,out}=Q_\text{V},
     \label{eq:vessel_mass_conservation}
 \end{equation}
and a pressure difference,
\begin{equation}
    P_\text{V,in} - P_\text{V,out} = \underbrace{\frac{8 \pi \mu l}{A^2}Q_\text{V} }_{\substack{\text{Term 1} \\ \text{(linear resistor)}}} + \underbrace{\frac{K_t \rho}{2A^2} \left( \frac{A}{A_\text{stenosis}} - 1\right)^2 Q_\text{V} |Q_\text{V} | }_{\substack{\text{Term 2} \\ \text{(stenosis resistor)}}} + 
\underbrace{\frac{\rho l}{A} \frac{d Q_\text{V}}{dt}}_{\substack{\text{Term 3} \\ \text{(inductor)}}} ,
    \label{eq:vessel_pressure_drop}
\end{equation}

are enforced.  Here $\mu$ and $\rho$ are the viscosity and density of blood, $A$ is the cross-sectional area of the vessel, and $K_t = 1.52$ is an empirically determined constant \cite{Steele2003InGrafts, Mirramezani2019ReducedArteries, Itu2013Non-invasiveMeasurements}.  At junctions, mass is similarly conserved,

 \begin{equation}
     Q_\text{J,in}- \sum_\text{J,out}Q_\text{J,out} = 0,
 \end{equation}
 
and pressure is assumed to be the same at the inlets and outlets \cite{Taylor-LaPole2023APatients, Olufsen1999StructuredArteries, Stergiopulos1992ComputerA, Reymond2009ValidationTree},

\begin{equation}
    P_\text{J,in} - P_\text{J,out} = 0 \quad \forall \text{ outlets}.
    \label{eq:constant_pressure_junction}
\end{equation}

\begin{figure}
    \centering
    \includegraphics[width=0.6\linewidth]{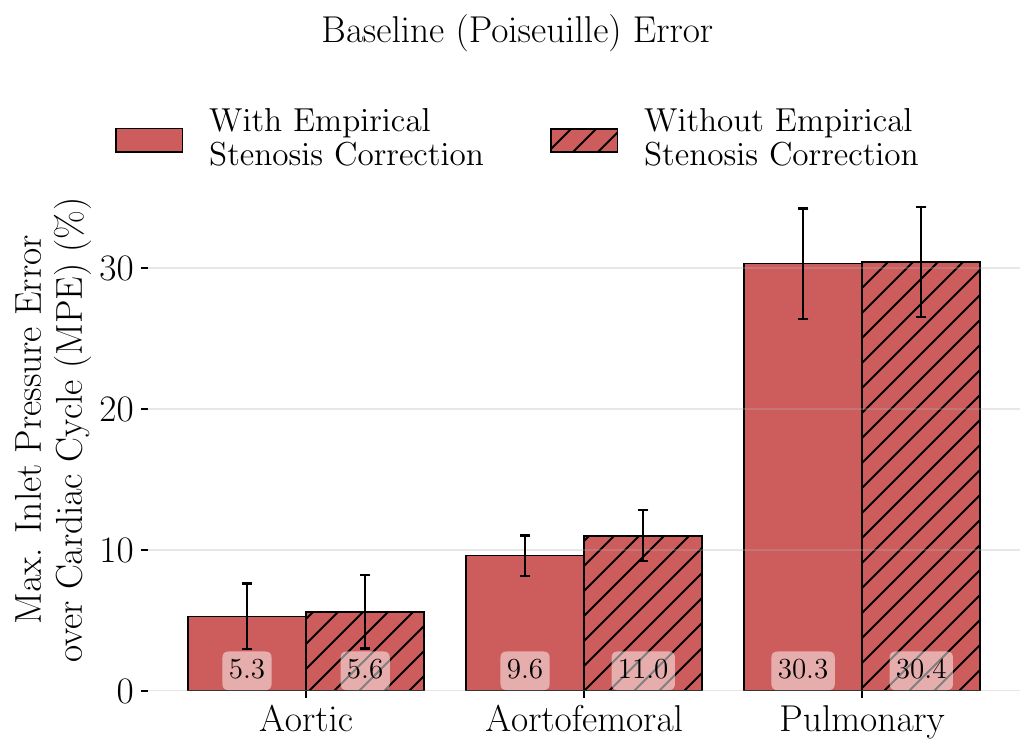}
    \caption{Maximum inlet percent error over cardiac cycle (MPE) using the baseline (Poiseuille) 0D model with and without the empirical stenosis correction (Term 2 of \Cref{eq:vessel_pressure_drop}). The mean and 95\% confidence interval are reported for each anatomical cohort.}
    \label{fig:empirical_stenosis}
\end{figure}

Terms 1 and 3 of \Cref{eq:vessel_pressure_drop} are arrived at by simplifying the Navier-Stokes equations under Poiseuille flow assumptions.  Term 1 captures viscous pressure losses and Term 3 describes inertial pressure differences induced by acceleration of the flow.  Term 2 is a small empirical correction capturing non-linear losses from contraction and expansion \cite{Steele2003InGrafts, Mirramezani2019ReducedArteries, Itu2013Non-invasiveMeasurements}.  We find that this correction yields a slight improvement in accuracy, as shown in \Cref{fig:empirical_stenosis}, and therefore include it in our standard 0D model. 

Notably, this set of equations resembles those describing an electric circuit, with pressure acting as voltage and flow rate as current  \cite{Ventre2020Reduced-OrderPathologies, Ghigo2017AArteries, Formaggia2010CardiovascularSystem}.  In this context, the $Q_\text{V,in}$ term in \Cref{eq:vessel_pressure_drop} acts as a linear resistor, the $Q_\text{V,in} |Q_\text{V,in} |$ term acts as a quadratic resistor, and the $\frac{d Q_\text{V,in}}{dt}$ term acts as an inductor. For this reason, 0D models are often referred to as electric circuit analogs.

\subsubsection{Modified Junction Handling}
\label{sec:junction_handling}

Recent work suggests that pressure differences over junctions are better modeled by a ``resistor-resistor-inductor" (RRI) model, 

\begin{equation}
    P_\text{J,in} - P_\text{J,out} = R_\text{lin} Q_\text{J,out} + R_\text{quad} Q_\text{J,out}| Q_\text{J,out} | + L \frac{dQ_\text{J,out}}{dt}.
    \label{eq:RRI_pressure_junction}
\end{equation}

than by the standard constant pressure assumption as in \Cref{eq:constant_pressure_junction} \cite{Rubio2025HybridDifferences}.  This is a similar structure to that of the 0D vessel model, where the pressure drop is computed as the sum of a linear resistor, $R_\text{lin}$, contribution capturing viscous effects proportional to the flow rate, a quadratic resistor, $R_\text{quad}$, contribution capturing non-linear pressure recovery and separation effects proportional to the square of the flow rate, and an inductor, $L$, contribution capturing inertial effects proportional to the time derivative of the flow rate.  We note that pressure recovery effects depend only on the speed of the flow, not on the direction, and would therefore be better represented by $\Delta P = R_\text{quad} Q^2$ in cases of flow reversal.  This work does not include cases of significant flow reversal, and preliminary studies do not reveal a significant change in results between the even and odd quadratic forms, so we retain the form $\Delta P = R_\text{quad} |Q| Q$ to remain consistent with existing work in 0D modeling.  In future work featuring large flow reversals, this should be reconsidered.   

Notably, while analytical formulas for $R_\text{lin}$, $R_\text{quad}$, $L$ exist for blood vessels, \Cref{eq:poiseuille_parameters},  such models do not exist for junctions.  Instead, we determine these parameters from the bifurcation geometry using a neural network.  (This will be discussed in further detail in \Cref{sec:neural_network}.)

In a study of the RRI bifurcation model applied to synthetic vascular trees \cite{Rubio2026Data-drivenModels}, a simplified resistor-inductor (RI) method which excludes the quadratic resistor was introduced,

\begin{equation}
    P_\text{J,in} - P_\text{J,out} = R_\text{lin} Q_\text{J,out} + L \frac{dQ_\text{J,out}}{dt}.
    \label{eq:RI_pressure_junction}
\end{equation}

While the RI model could not fit 3D data as accurately as the RRI model, it still reduced 0D model error significantly, and its simplified form reduced computational complexity for the forward 0D simulation compared to the RRI model.  Furthermore, the simpler RI coefficients likely have less complex relationships to the vascular geometry, and may prove easier to ``learn" for a neural network, particularly with limited training data.  

In \cite{Rubio2026Data-drivenModels}, it was found that the main improvements in accuracy resulting from use of the RRI model rather than the RI model were made in the most proximal bifurcations (closest to the inlet).  To retain these benefits without introducing numerical complexity that necessitates alternative solution strategies, as in \cite{Rubio2026Data-drivenModels}, when we indicate that we are using the RRI model, we only apply the RRI model to the first (most proximal) bifurcation and vessel.  In the more distal elements, we use the RI model.  When we indicate that we are using the RI model, we use the RI model for all elements.

\subsubsection{Modified Vessel Handling}
In this work, we also test an extension of the RRI methodology to vessels.  Having seen that the RRI model with learned coefficients yielded improved accuracy on synthetic trees when applied to bifurcations \cite{Rubio2026Data-drivenModels}, we hypothesize that further reductions to 0D model error can be achieved by using the RRI model with learned coefficients for vessels.  The standard approach models pressure drops over vessels using a similar form (linear resistance, quadratic resistance, and inductance) \Cref{eq:vessel_pressure_drop}, but uses different coefficients:

\begin{equation}
    R_\text{Poiseuille} = \frac{8 \pi \mu l}{A^2}, \quad R_\text{stenosis} =  \frac{K_t \rho}{2A^2} \left( \frac{A}{A_\text{stenosis}} - 1\right)^2, \quad L = \frac{\rho l}{A},
    \label{eq:poiseuille_parameters}
\end{equation}

derived from  Poiseuille flow assumptions and an empirical stenosis model, shown to improve agreement with 3D simulations in vessels featuring a constriction \cite{Steele2003InGrafts, Mirramezani2019ReducedArteries, Itu2013Non-invasiveMeasurements}.  Since the flow profiles in vascular anatomies can deviate significantly from Poiseuille flow, these coefficients may introduce error into the 0D model's characterization of vessel hemodynamics, contributing to overall error.  We hypothesize that further error reduction can be achieved by replacing these coefficients with those learned from the 3D data.  That is, we replace \Cref{eq:vessel_pressure_drop} with an RRI vessel model,

\begin{equation}
    P_\text{V,in} - P_\text{V,out} = R_\text{lin} Q_\text{V} + R_\text{quad} |Q_\text{V}| Q_\text{V} + L \frac{dQ_\text{V}}{dt},
    \label{eq:RRI_pressure_vessel}
\end{equation}

or RI vessel model,

\begin{equation}
    P_\text{V,in} - P_\text{V,out} = R_\text{lin} Q_\text{V} + L \frac{dQ_\text{V}}{dt},
    \label{eq:RI_pressure_vessel}
\end{equation}

where $R_\text{lin}$, $R_\text{quad}$, $L$ are predicted from the vessel geometry using neural networks.

\subsection{Neural Networks}
\label{sec:neural_network}
In this section, we discuss the neural networks used to predict $R_\text{lin}$, $R_\text{quad}$, and $L$ for vessels and bifurcations from their geometric features.  We use a neural network of the form

\begin{align*}
    y_0 & = \text{ReLU}(W_{0} \mathcal{G} + b_{0}) \\
    y_i & = \text{ReLU}(W_{i} y_{i-1} + b_{i}) \;\;\; \forall i = 1...n-1 \\
   \mathcal{O} & = W_{n} y_{n-1} + b_{n} ,
\end{align*}

to predict each of the three outputs $\mathcal{O} = R_\text{lin}, R_\text{quad}, \text{or } L$ from a geometric input vector $\mathcal{G}$.  Since we use separate neural networks to predict the parameters for junctions and vessels, there are six networks total. The number of layers, ($n-1$), ranges from 2-4 and the layer width, (the size of $W$ and $b$), ranges from 10-20 neurons for each hidden layer.  Specifics for each network are reported in \Cref{sec:app_nn_architecture}. We note that, at inference only, the neural network outputs are restricted to lie within the range observed in the training set.  The weights, $W$, and biases, $b$, for each network are optimized to minimize a mean-squared error (MSE) loss,

\begin{equation}
        \text{Loss} = \left \lVert \mathcal{O}_\text{pred} - \mathcal{O}_\text{true} \right \rVert_2
        \label{eq:mse_loss}
\end{equation}

We also consider an alternative loss that weighs data from more proximal (closer to the vasculature inlet) elements (junctions or vessels) more heavily to correct for over-representation of distal elements in the training data.  Generation is represented by the generation number $\gamma$, which is equal to the number of bifurcations between the element and the vasculature inlet.  We define this generation-weighted loss as

\begin{equation}
        \text{Loss} = \frac{1}{2^\gamma} \left \lVert \mathcal{O}_\text{pred} - \mathcal{O}_\text{true} \right \rVert_2.
        \label{eq:prox_loss}
\end{equation}

We use an adaptive moment estimator (Adam) optimizer from the \texttt{optax} library to identify the weights, $W$, and biases, $b$.  We implement the neural networks using \texttt{JAX} for accelerated training.

\subsubsection{Neural Network Inputs - Geometric Features}
The inputs to the neural networks are vectors describing the bifurcation or vessel geometry.  The junction input vector, $\mathcal{G}_J$ contains all the features listed in \Cref{tab:geo_features} except the vessel feature ``\textit{is inlet}", and the vessel input vector, $\mathcal{G}_V$ contains all except the flow split $\phi$.  We consider these geometric parameters because they can be computed from the geometry without running a 3D simulation.

\begin{table}[htbp]
\renewcommand{\arraystretch}{1.8} 
\begin{tabular}{>{\raggedright\arraybackslash}p{2.3cm} >{\raggedright\arraybackslash}p{2cm} >{\raggedright\arraybackslash}p{10cm}}
\hline
  Inlet radius:  & $\displaystyle r_\text{in}$, & \textit{Maximum inscribed sphere radius (MISR) at inlet.} \\
  Outlet radius: & $\displaystyle r_\text{out},r_\text{out}^{-2},r_\text{out}^{-4}$ & \textit{MISR at outlet.}\\
  Minimum radius: & $\displaystyle r_\text{min}$, & \textit{Smallest MISR on path from inlet to outlet.}\\
  Maximum radius: & $\displaystyle r_\text{max}$, & \textit{Largest MISR on path from inlet to outlet.}\\
  \multirow{3}{*}{\shortstack[l]{Radius \\ ratios: }}  & $\displaystyle r_\text{out}^* = \frac{r_\text{out}}{r_\text{in}}$ & \textit{Ratio of outlet MISR to inlet MISR. }\\
  & $\displaystyle r_\text{min}^* = \frac{r_\text{min}}{r_\text{out}}$ & \textit{Ratio of minimum MISR to outlet MISR.} \\
  & $\displaystyle r_\text{max}^* = \frac{r_\text{max}}{r_\text{out}}$ & \textit{Ratio of maximum MISR to outlet MISR.}  \\
  Length: & $l$ & \textit{Centerline path length from inlet to outlet.} \\
  Length ratio: & $\displaystyle l^* = \frac{l}{r_\text{inlet}}$ & \textit{Ratio of path length to inlet MISR.} \\
  Tortuosity: & $\tau = \frac{l}{d}$ & \textit{Ratio of path length $l$ to straight-line distance from inlet to outlet,} $d$. \\
  Angle: & $\theta$ & \textit{Angular difference between centerline tangent at inlet and outlet.} \\
  Is inlet: & $\delta$ & \textit{1 if the vessel is connected to an inlet boundary condition, 0 otherwise.  (Vessels only)} \\
  Flow ratio: & $\displaystyle \phi = \frac{Q_\text{inlet}^{\text{avg}}}{Q_\text{outlet}^{\text{avg}}} $ & \textit{Ratio of inlet flow to outlet flow  (as predicted by baseline 0D simulation), averaged over the cardiac cycle.  (Junctions only)} \\
  \multirow{3}{*}{\shortstack[l]{Absorbed \\ parameters: }}  & $\displaystyle R_\text{Poiseuille}^\text{absorbed}$ & \multirow{3}{*}{\shortstack[l]{\textit{Baseline (Poiseuille) 0D parameters associated with portions of  } \\ \textit{outlet vessels that are absorbed into the junction during }\\ \textit{entrance-length adjustment. For vessels, baseline 0D parameters } \\ \textit{minus the portion absorbed into a preceding junction.}}} \\
   & $\displaystyle R_\text{stenosis}^\text{absorbed} $& \; \\
   & $\displaystyle L^\text{absorbed}$ & \; \\
     \multirow{3}{*}{\shortstack[l]{Calculated \\ parameters: }}  & $\displaystyle R_\text{Poiseuille}^\text{calculated}$ & \multirow{3}{*}{\shortstack[l]{\textit{Poiseuille parameters calculated using \Cref{eq:poiseuille_parameters} using length} $l$ \\ \textit{and radius} $r_\text{outlet}.$ \textit{This differs from the absorbed parameters} \\ \textit{primarily in that} $l$ \textit{includes segment of the centerline inside}\\ \textit{the junction region.}}} \\
   & $\displaystyle R_\text{stenosis}^\text{calculated} $ &  \\
   &  $\displaystyle L^\text{calculated} $ &  \\
   \hline
\end{tabular}
\caption{Geometric features used as input to the neural networks used to predict lumped parameters $R_\text{lin}$, $R_\text{quad}$, $L$, for vessel and junctions.}
\label{tab:geo_features}
\end{table}

Most of these values are extracted from the 3D vasculature using centerlines generated with \texttt{SimVascular}\footnote{\url{https://simvascular.github.io/}}, an open-source software supporting patient-specific cardiovascular modeling from medical images.  Features are extracted at centerline points associated with the inlets and outlets of the vessel or bifurcation.  Areas are found by integrating over the contiguous cross-section normal to the centerline tangent at a given centerline point.  Radii are defined using the maximum inscribed sphere radius, which refers to the maximum radius of a sphere centered at a given centerline point and completely contained in the vessel.  Angles are found by comparing centerline tangents at inlets and outlets.  Path lengths refer to the length of the centerline segment between inlet and outlet, computed by summing distances between consecutive points.  The flow ratio, $\phi$, is computed for junctions based on the flows predicted by a standard 0D simulation.

\subsubsection{Neural Network Outputs - Optimal 0D Parameters}
The neural networks predict outputs $R_\text{lin}$, $R_\text{quad}$, $L$.  The ground truth values of these parameters are those that minimize the differences between the 0D solution and high-fidelity 3D finite-element simulation.  To find these, we first project the 3D solution into 1D space by integrating time-resolved values of velocity, $\vec{u}(t)$, and pressure, $p(t)$, over the vessel cross-sectional area, 

\begin{equation}
    P(t) = \frac{\int_A p(t) \; dA}{\int_A  dA}, \qquad Q(t) = \int_A \vec{u}(t)\cdot \vec{n} \; dA,
\end{equation}

yielding bulk solution values, flow rate $Q(t)$ and average pressure $P(t)$ for each point along the centerlines of the vasculature.  To arrive at the 0D representation of the 3D solution, we extract these bulk solution values, $Q(t)$ and $P(t)$ at the centerline points corresponding to 0D nodes, that is, at the inlets/outlets of the 0D vessels and junctions.  

To find the optimal $R_\text{lin}$, $R_\text{quad}$, and $L$ for the RRI model, we use an approach similar to that in \cite{Rubio2026Data-drivenModels, Rubio2025HybridDifferences}, a least squares fit to the data obtained from 3D simulation:
\begin{gather*}
\Delta P(t) = R_{\text{lin}} Q(t) + R_{\text{quad}} Q(t) |Q(t)| + L \dot{Q}(t), \\
\forall \; t \in \text{simulation timesteps where } Q(t) > 10^{-12}.
\end{gather*}

When the RI model is considered, the quadratic resistor is forced to have a value of 0, and the least squares problem becomes

\begin{gather*}
\Delta P(t) = R_{\text{lin}} Q(t) +  L \dot{Q}(t), \\
\forall \; t \in \text{simulation timesteps where } Q(t) > 10^{-12}.
\end{gather*}

To ensure physical consistency, each coefficient is forced to have the appropriate sign.  While uncommon, if the standard least-squares problem does not yield appropriately signed coefficients, \texttt{scipy}'s Trust Region Reflective algorithm is used to find the nearest correctly signed solution.  Inductance is always assumed to be positive, as it represents the fluid's resistance to change in speed.  In the RRI model, the quadratic resistance may be positive (due to non-linear losses or flow acceleration over a bifurcation) or negative (due to transformation of kinetic energy to pressure energy as fluid decelerates over a bifurcation).  Linear resistance is assumed to be positive (as it captures only viscous pressure losses).  In the RI model, however, the linear resistance is allowed to be either positive or negative as it captures all steady behavior.  Note that these constraints are rarely active, the 3D data almost always produces coefficients of the expected sign.

\subsection{Geometric Preprocessing}
\label{sec:geometric_preprocessing}
Representing a 3D vasculature as a 0D model requires a discretization of the geometry into vessels and junctions.  In the \texttt{SimVascular ROM Simulation Tool}\footnote{\url{https://simvascular.github.io/documentation/rom_simulation.html\#tool}}, described in \cite{Pfaller2022AutomatedFlow}, this is done using the 3D geometry and vessel centerlines according to two criteria: 

\begin{enumerate}
    \item If the 3D mesh points included in a centerline-normal, contiguous slice of a 3D anatomy have different values of nearest centerline branch, that centerline point is part of a junction.
    \item If another centerline is included in a centerline-normal, contiguous slice of a 3D anatomy, that centerline point is part of a junction.
\end{enumerate}

This is a robust method for producing vessel-junction discretizations for standard 0D modeling, but to apply our hybrid methodology, we propose two post-processing steps to slightly modify the vessel-junction discretization.

\subsubsection{Bifurcation Splitting}
The standard vessel-junction discretization yields junctions with one inlet and an arbitrary number of outlets.  When junctions are relatively close together, for example, in pulmonary anatomies, this can sometimes yield large junctions with ten or more outlets, which would be better represented as a series of bifurcations.  The neural networks used to predict junction RRI parameters in this work require a geometric description vector, $\mathcal{G}_J$, of a fixed size as input.  Although in this work we only use features describing the inlet and outlet of interest, this feature vector might be extended to include features describing the other junction outlets, as in \cite{Rubio2025HybridDifferences, Rubio2026Data-drivenModels}.  If the number of additional outlets is allowed to vary, as in the n-outlet junction definition, the size of $\mathcal{G}_J$ will also vary, necessitating either a more complex machine-learning model form or a different neural network for each junction type with different numbers of outlets.

\begin{figure}[htbp]
    \centering
    \includegraphics[width=0.9\linewidth]{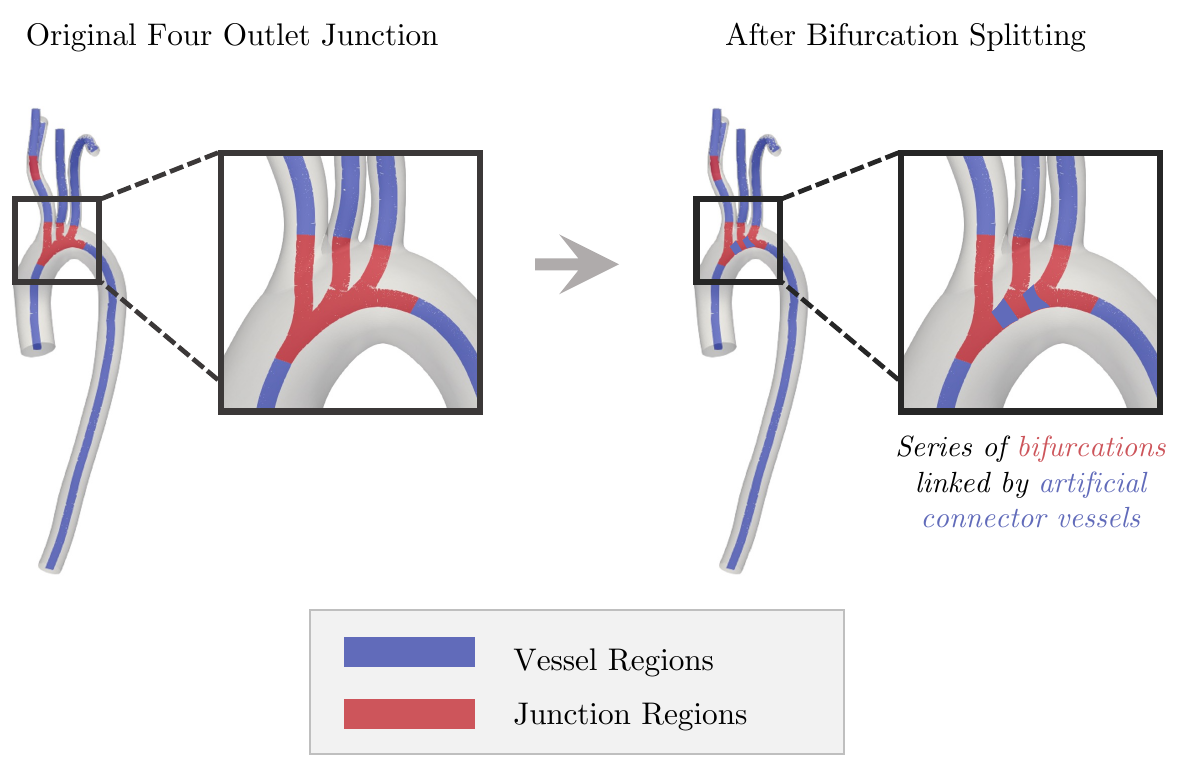}
    \caption{Definition of junction and vessel regions before and after applying the bifurcation splitting algorithm.  A four-outlet junction becomes a series of three bifurcations connected by artificial connector vessels.  These artificial connector vessels have zero length and do not affect hemodynamics, but are shown here with non-zero size for illustration purposes.}
    \label{fig:bifurcation_splitting}
\end{figure}

To avoid these issues and more appropriately describe the anatomy, we apply a post-processing step to transform all junctions with three or more outlets into a series of bifurcations (two-outlet junctions).  Upon encountering a multi-outlet junction, we sort the outlets based on the in-junction path length of the centerline connecting the junction inlet to the outlet, $l_j$.  Then, we create a bifurcation with an inlet corresponding to the original junction inlet, one outlet connecting to the junction outlet with the shortest path length $l$ between inlet and outlet, and one outlet connecting to an artificial ``connector vessel 0".  We repeat this process, creating another bifurcation with ``connector vessel 0" as inlet, the outlet with next shortest in-junction path length as one outlet, and another ``connector vessel 1" as the second outlet.  We continue in this fashion until only the two outlets with the longest in-junction paths remain.  A final bifurcation is added with the last connector vessel as an inlet, and the two outlets of the original junction with the longest in-junction paths as outlets.  We show a visualization of this in \Cref{fig:bifurcation_splitting} and provide a description of the splitting algorithm in \Cref{alg:bifurcation_splitting}.

\begin{algorithm}
\label{alg:bifurcation_splitting}
\begin{algorithmic}
\Require $l_{j,i} \leq l_{j, i+1} \quad \forall i \in [0, n-1]$ \Comment{Sort $n$ outlets in order of junction length $l_j$}

\If{$n > 2$}
\Comment{Check if multi-outlet junction}
    \For{$i = 0:n-2$ } \Comment{Loop over $n-2$ intermediate bifurcations}
    \State Create bifurcation $i$ 
    \State Create connector vessel $i$
    \If{$i = 0$}
        \State Set bifurcation $i$ inlet to be the original junction inlet
    \Else
        \State Set bifurcation $i$ inlet to be connector $i-1$
    \EndIf
    \State Set first bifurcation $i$ outlet to be outlet $i$ of the original junction
    \State Set second bifurcation $i$ outlet to be connector $i$
    \EndFor
    
    \State Create bifurcation $n-2$ 
    \Comment{Final bifurcation}
    \State Set bifurcation $n-2$ inlet to be connector $n-3$
    \State Set first bifurcation $i$ outlet to be outlet $n-2$ of the original junction
    \State Set second bifurcation $i$ outlet to be connector $n-1$ of the original junction
\Else
\Comment{Two-outlet junction}
    \State Create bifurcation 0
    \State Set bifurcation 0 inlet to be the original junction inlet
    \State Set first bifurcation 0 outlet to be outlet 0 of the original junction
    \State Set second bifurcation 0 outlet to be connector 1 of the original junction
\EndIf
\end{algorithmic}
\caption{Procedure to convert a multi-outlet junction into a series of bifurcations.}
\end{algorithm}

\subsubsection{Entrance Length Adjustment}

\begin{figure}[htbp]
    \centering
    \includegraphics[width=0.90\linewidth]{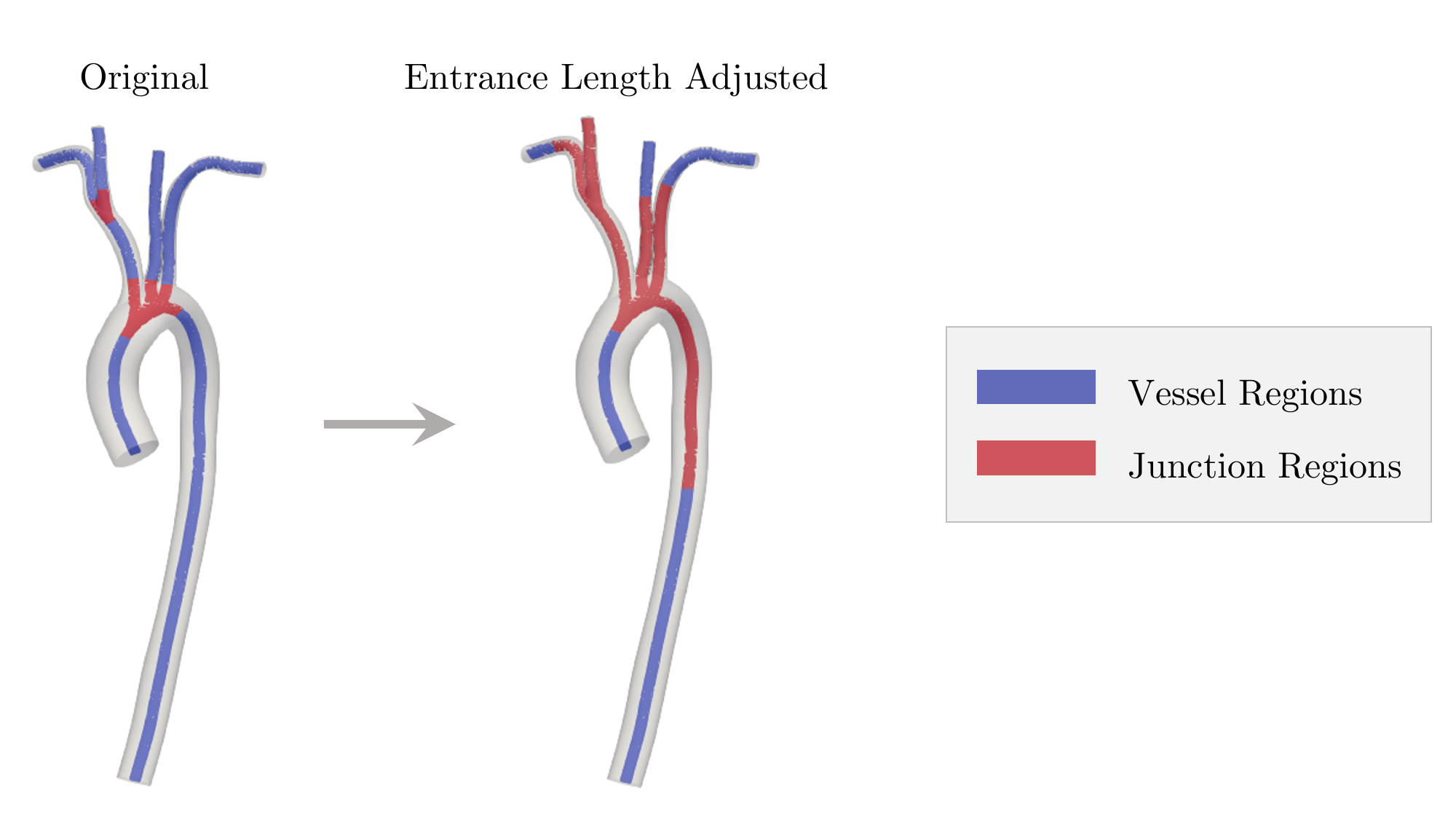}
    \caption{Discretization of vasculature into junctions and vessels before and after entrance length adjustment.  Junction boundaries are extended into the outlet branches by a pseudo-entrance length.  In some cases, where the outlet vessel is shorter than the pseudo-entrance length, the outlet vessel is completely absorbed into the junction.}
    \label{fig:entrance_length_adjustment}
\end{figure}

\texttt{SimVascular}'s vessel - junction discretization algorithm draws the boundaries of junctions close to the site of the split, as shown in \Cref{fig:entrance_length_adjustment}.  The flow in this region is subject to significant localized 3D effects (e.g. a high-pressure zone in the impingement region, recirculation, etc.), that are irrelevant in a 0D modeling context, but create significant ``noise" in the measured 3D $\Delta P$, and therefore in the $R_\text{lin}$, $R_\text{quad}$, and $L$ values that are fit to reproduce the 3D $\Delta P$.  In turn, this makes the task of learning the $R_\text{lin}$, $R_\text{quad}$, and $L$ values from the geometry more challenging.

To avoid this issue, we introduce another post-processing step where we extend the boundaries of the junctions into the outlet vessels by a ``pseudo entrance length", $L_e$.  At this distance, the flow has developed in the vessel and is less subject to such noise.  A true entrance length depends on the Reynolds number of the flow, but these vasculatures experience a wide range of Reynolds numbers over the cardiac cycle and we must choose a single vessel-junction discretization to use over the full simulation.  One approach may be to choose an entrance length based on the maximum Reynold's number experienced over the cardiac cycle (determined by the inflow boundary condition).  In some cases, however, the inlet boundary condition may not be available at the time of the 0D model generation.  For this reason, we choose a flow-independent surrogate for entrance length, $L_e = 10 r_{outlet}$, where $r_{outlet}$ is the radius of the outlet vessel.  Within the Reynolds numbers we consider, we find this to be a conservative choice of pseudo entrance length where the flow is free of localized splitting effects.  We extend the boundaries of each junction outlet a distance of $L_e$ into the outlet vessel.  If the outlet vessel is shorter than $L_e$, the entire vessel is absorbed into the junction, and an artificial, zero-length connector vessel is created to couple the junction outlet to the downstream junction or boundary condition.

\subsection{Pipeline Overview}
\begin{figure}[htbp]
    \centering
    \includegraphics[width=1\linewidth]{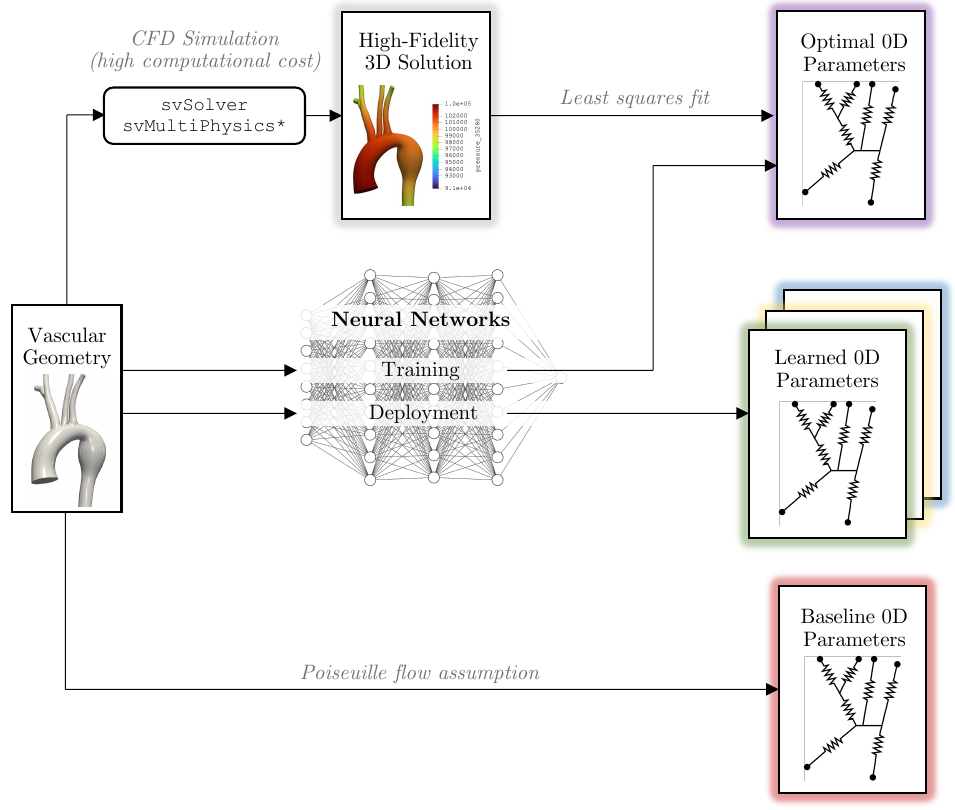}
    \caption{Overview of different 0D electric circuit analog modalities.  0D parameters may be generated by fitting to high-fidelity 3D simulation, neural network prediction, or traditional Poiseuille assumption formulas.  The 0D model parameter sets are colored to match their associated forward-simulation accuracies shown in \Cref{fig:cv_max_percent_error_bars}.  Neural networks learn the relationship between training geometries and optimal parameters and then are deployed to predict 0D parameters for unseen test geometries, avoiding the costly 3D CFD simulation step.}
    \label{fig:pipeline_schematic}
\end{figure}

In this section, we provide a summary of the hybrid data-driven and physics-based reduced-order modeling framework proposed in this work, a high-level representation of which is shown in \Cref{fig:pipeline_schematic}.  We create an electric circuit analog of a 3D vascular geometry using existing centerline-based tools in \texttt{SimVascular} and the bifurcation-splitting and entrance-length adjustment pre-processing steps introduced above. Then, we use neural networks to predict the parameters $R_\text{lin}$, $R_\text{quad}$, and $L$ that best represent each vessel and junction in the vasculature from their geometries.  The neural networks are trained to minimize the difference between the parameters they predict and the optimal parameters, which are determined from high-fidelity 3D CFD simulation using a least squares fit.

As a control, we also consider a traditional, purely physics and heuristic-based 0D analog, where the pre-processing steps are skipped, and the parameters $R_\text{lin}$, $R_\text{quad}$, and $L$ are determined from \Cref{eq:poiseuille_parameters}, simplifications of the Navier-Stokes equations based on a Poiseuille flow assumption.  Tools for generating this type of standard 0D model are available as part of the open-source cardiovascular modeling suite \texttt{SimVascular ROM Simulation Tool}.  We refer to this modality of electric circuit analog as ``standard", ``baseline" or ``Poiseuille".

Once the 0D parameters have been chosen, a forward 0D simulation can be run to yield the 0D prediction of flow rate and pressure in the vasculature over a cardiac cycle.  The 0D simulation runtime using \texttt{svZeroDSolver} \footnote{\url{https://github.com/SimVascular/svZeroDSolver}} is <2 seconds for the most complex pulmonary anatomies and <1 second for the simpler geometries.  This 0D solution is the primary output of interest, and the metrics we consider measure how much the 0D solution differs from the high-fidelity 3D solution, which we take as the ground truth.

We compare the results of the forward 0D simulations for 0D models that use the baseline (Poiseuille) parameters, neural network-predicted junction parameters with baseline vessel parameters, baseline junction parameters with neural network-predicted vessel parameters, neural network-predicted junction and vessel parameters, and the optimal (fit to 3D simulation) parameters for both vessels and junctions.

We also test the effects of adopting the novel numerical methodologies we propose: entrance-length adjustment and generation-weighted neural-network optimization loss \Cref{eq:prox_loss}.  We further investigate the effect of including the quadratic resistor, $R_\text{quad}$, as in the full RRI model (\Cref{eq:RRI_pressure_junction}, \Cref{eq:RRI_pressure_vessel}), or using the simpler RI model (\Cref{eq:RI_pressure_junction}, \Cref{eq:RI_pressure_vessel}).

\subsection{Data}
\begin{figure}[htbp]
    \centering
    \includegraphics[width=1\linewidth]{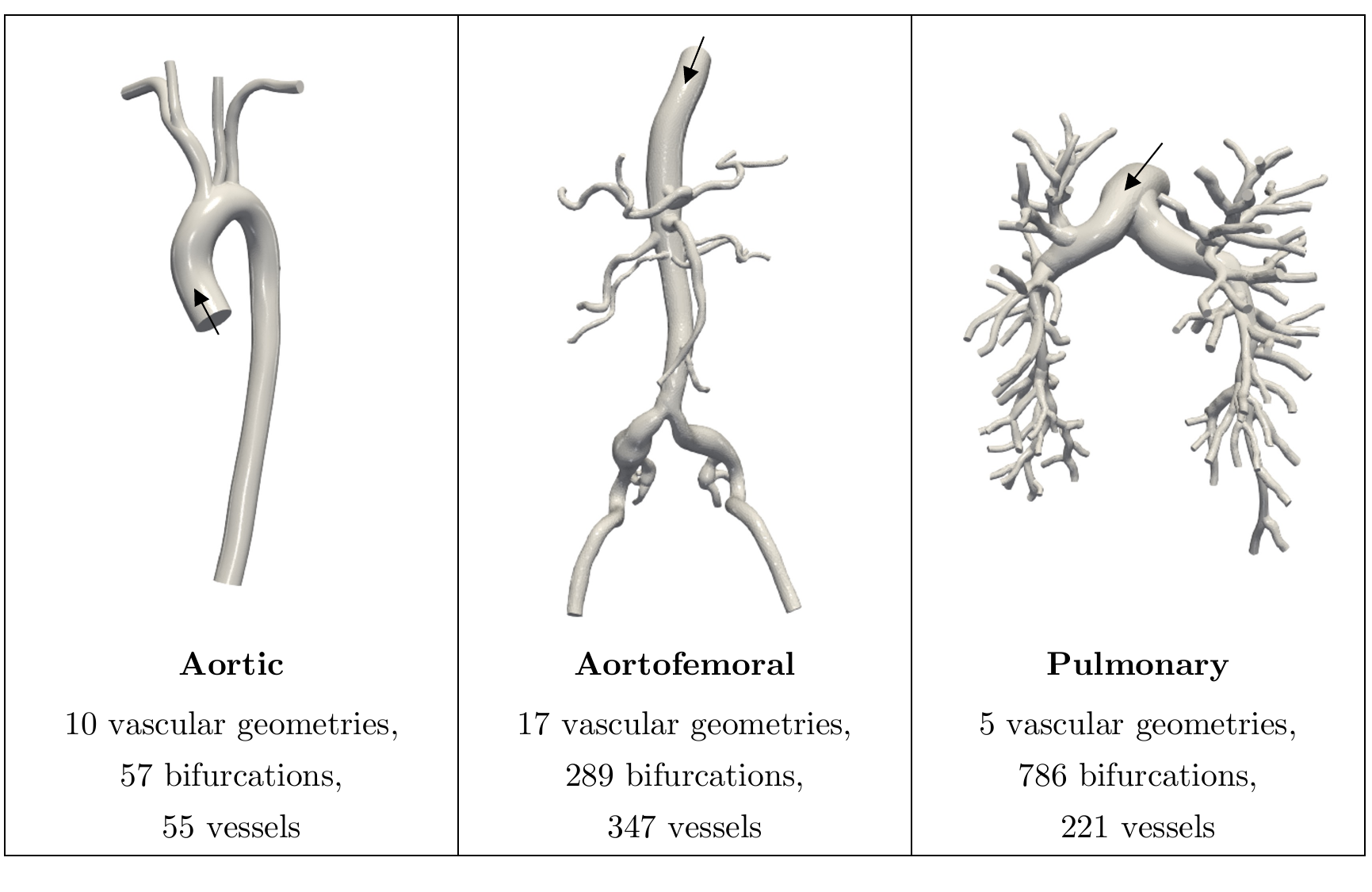}
    \caption{Representative geometry for each of the three anatomy types (aortic, aortofemoral, and pulmonary) on which we test the effects of using neural-net predicted 0D parameters.  Arrows mark the vasculature inlets where we measure pressure errors.  For each cohort, we report the number of bifurcation elements and vessel elements in the dataset.  Note that the number of bifurcations here refers to the number of bifurcation inlet - outlet pairs, and that we do not count artificial connector vessels or bifurcation outlets leading to connector vessels.  A vascular geometry may contain more bifurcations than vessels in cases where many vessels have been fully absorbed into the junction regions during the entrance length adjustment preprocessing step.}
    \label{fig:aorta_cohort}
\end{figure}

We consider four cohorts of patient-specific vasculatures with 3D simulations sourced from the dataset created in \cite{Pfaller2022AutomatedFlow} and available at the \texttt{Vascular Model Repository} \footnote{\url{https://www.vascularmodel.com/index.html}}.  The four cohorts are: a set of 10 aortic anatomies, a set of 17 aortofemoral anatomies, a set of 5 pulmonary anatomies, and a mixed ``all" set containing all the anatomies.  Examples of aortic, aortofemoral, and pulmonary anatomies are shown in \Cref{fig:aorta_cohort}.  Note that vasculatures from the same anatomy type may not necessarily contain all the same vessels, for example, some aortic anatomies include the brachiocephalic bifurcation in the top left of the left panel of \Cref{fig:aorta_cohort}, while others do not.  We report the total number of bifurcation and vascular elements in each cohort in \Cref{fig:aorta_cohort}.

The vascular geometries were constructed using \texttt{SimVascular} and the 3D simulations were run using \texttt{svSolver}\footnote{\url{https://github.com/SimVascular/svSolver}}. Simulations for these anatomies were run with rigid walls, Windkessel or resistance outlet boundary conditions, and a prescribed, patient-specific, time-varying parabolic inflow, as shown in \Cref{fig:pressure_in_time}.

As we have relatively little data, we use five-fold cross-validation to add statistical significance to our evaluation of the performance of the 0D models with neural-network predicted parameters.  In each of five trials, $10\%$ of the geometries are set aside for testing, while the bifurcation and vessel elements from the other $90\%$ of the geometries are used to train the neural networks to predict the 0D circuit values $R_\text{lin}$, $R_\text{quad}$, and $L$ for junctions and vessels.  These networks are then used to predict $R_\text{lin}$, $R_\text{quad}$, and $L$ for the bifurcations and vessels in the unseen test geometries.  Together, the five trials provide a balanced characterization of the accuracy of the 0D model with neural-network-predicted parameters compared to the accuracy of the baseline 0D model with Poiseuille assumption-derived parameters.

\section{Results}

\begin{figure}[htbp]
    \centering
    \includegraphics[width=1 \linewidth]{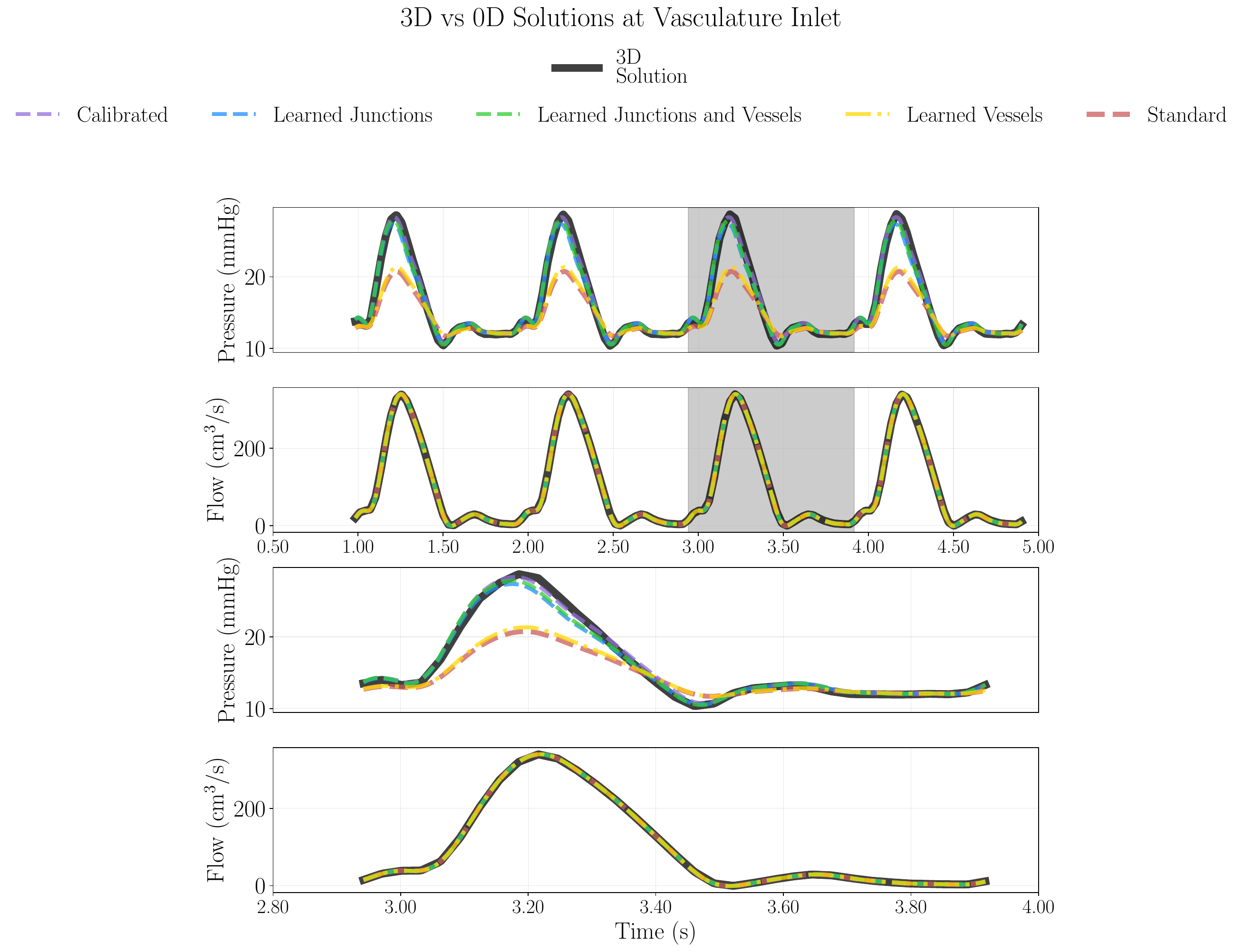}
    \caption{Inlet pressure and flow rate waveforms as predicted by 3D simulation and 0D simulation using standard Poiseuille parameters, learned parameters, and optimal parameters fit to the 3D solution.  MPE is reported for each 0D solution. The bottom two panels show a zoomed-in view of the shaded region in the top panels, representing one cardiac cycle.  The data shown is for a pulmonary anatomy (0079\_H\_PULM\_H).  The learned parameters are predicted by neural networks that have been trained on the remainder of the pulmonary cohort in one of the cross-validation trials.}
    \label{fig:pressure_in_time}
\end{figure}

To quantify the error associated with a 0D model, we consider the pressure at the vasculature inlet. Since we apply a flow boundary condition at the inlet and Windkessel or resistance boundary conditions at the outlet caps, flow rate error is expected to be zero at the inlet, while pressure error is expected to be low at outlets, increase for more proximal parts of the anatomy, and be large at the inlet, as is the case in \Cref{fig:pressure_in_time}.  We use a relative error metric (computed with respect to the high-fidelity 3D solution) at the point in the cardiac cycle where the absolute error is the greatest.  Hereafter, we refer to this metric, given by 
\begin{equation}
       \text{MPE} = \frac{|P_\text{inlet}(t^*)^{\text{0D}}-P_\text{inlet}^{\text{3D}}(t^*)|}{P_\text{inlet}^{\text{3D}}(t^*)} \times 100\%,
\end{equation}

\begin{equation*}
    \text{where } t^* = \text{argmax} (|P_\text{inlet}(t^*)^{\text{0D}}-P_\text{inlet}^{\text{3D}}(t^*)|), \; {t\in \text{cycle timesteps}},
\end{equation*}

as MPE (maximum percent error).  Typically, as shown in \Cref{fig:pressure_in_time}, this occurs at systole, the time in the cardiac cycle when the heart is contracting and flow rate is at its maximum.

\begin{figure}[htbp]
    \centering
    \includegraphics[width=1\linewidth]{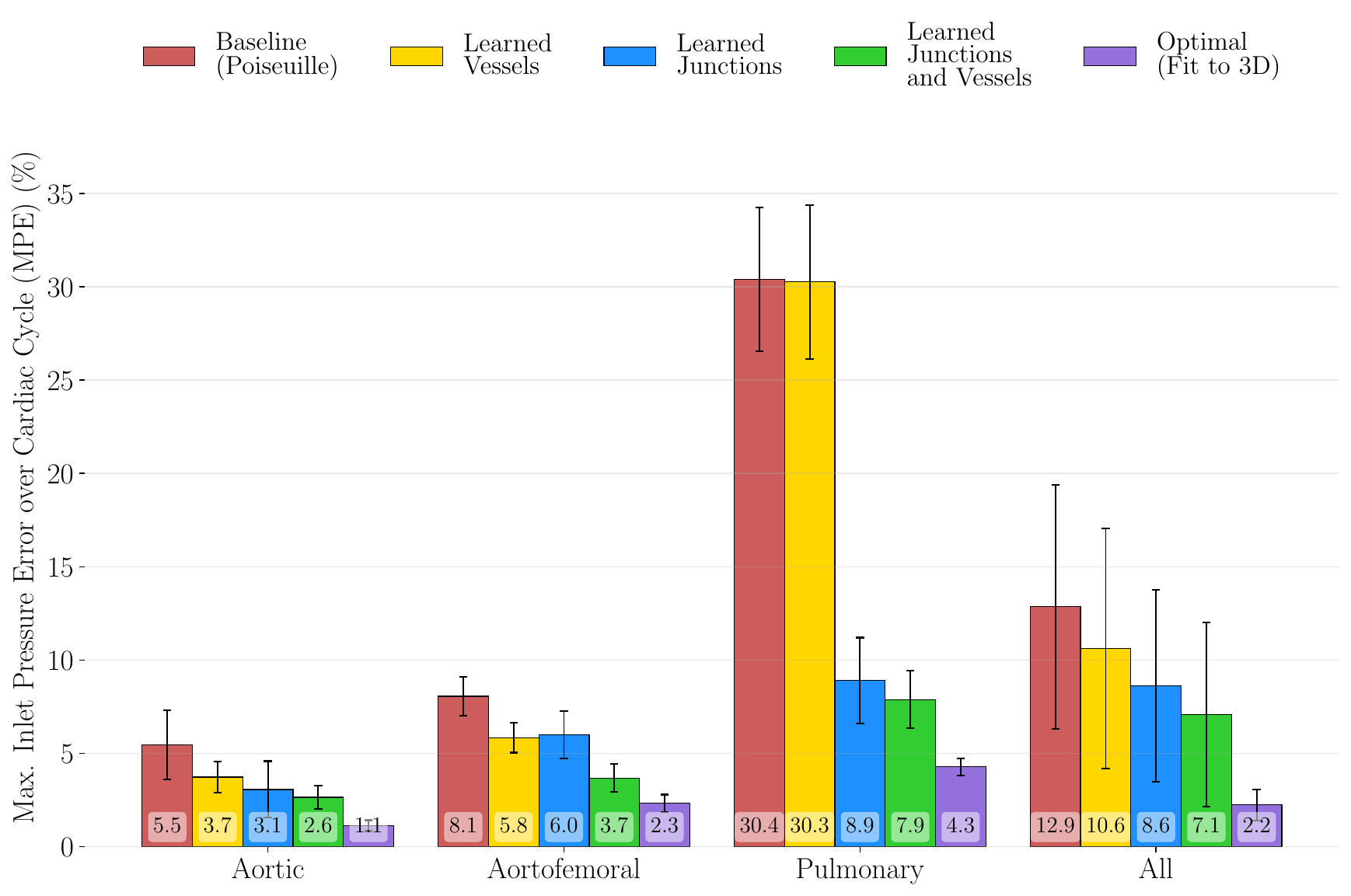}
    \caption{0D MPE average and 95\% confidence intervals computed over five cross-validation trials.  Bars for 0D models using baseline parameters, using neural network predicted (learned) parameters for vessels only, junctions only, and both vessels and junctions, and using optimal parameters obtained by fitting to 3D simulation are shown in different colors.  Results for each of the anatomical cohorts (aortic, aortofemoral, pulmonary, and all) are shown.}
    \label{fig:cv_max_percent_error_bars}
\end{figure}

In the five-fold cross-validation studies, shown in \Cref{fig:cv_max_percent_error_bars}, using learned parameters for vessels or junctions improves 0D model accuracy over the baseline parameters for all cohorts.  Learning both vessel and junction parameters further reduces the 0D error for all cohorts.  In the aortic cohort, learning the full 0D model parameters reduces the average MPE from $5.5\%$ to $2.6\%$, in the aortofemoral cohort from $8.1\%$ to $3.7\%$.  Learning the 0D model parameters makes the biggest difference in the pulmonary cohort, where MPE is reduced from $30.4\%$ to $7.9\%$. In the cohort including all anatomical types, MPE is reduced from $12.8\%$ to $7.1\%$.

\begin{figure}[htbp]
    \centering
    \includegraphics[width=1\linewidth]{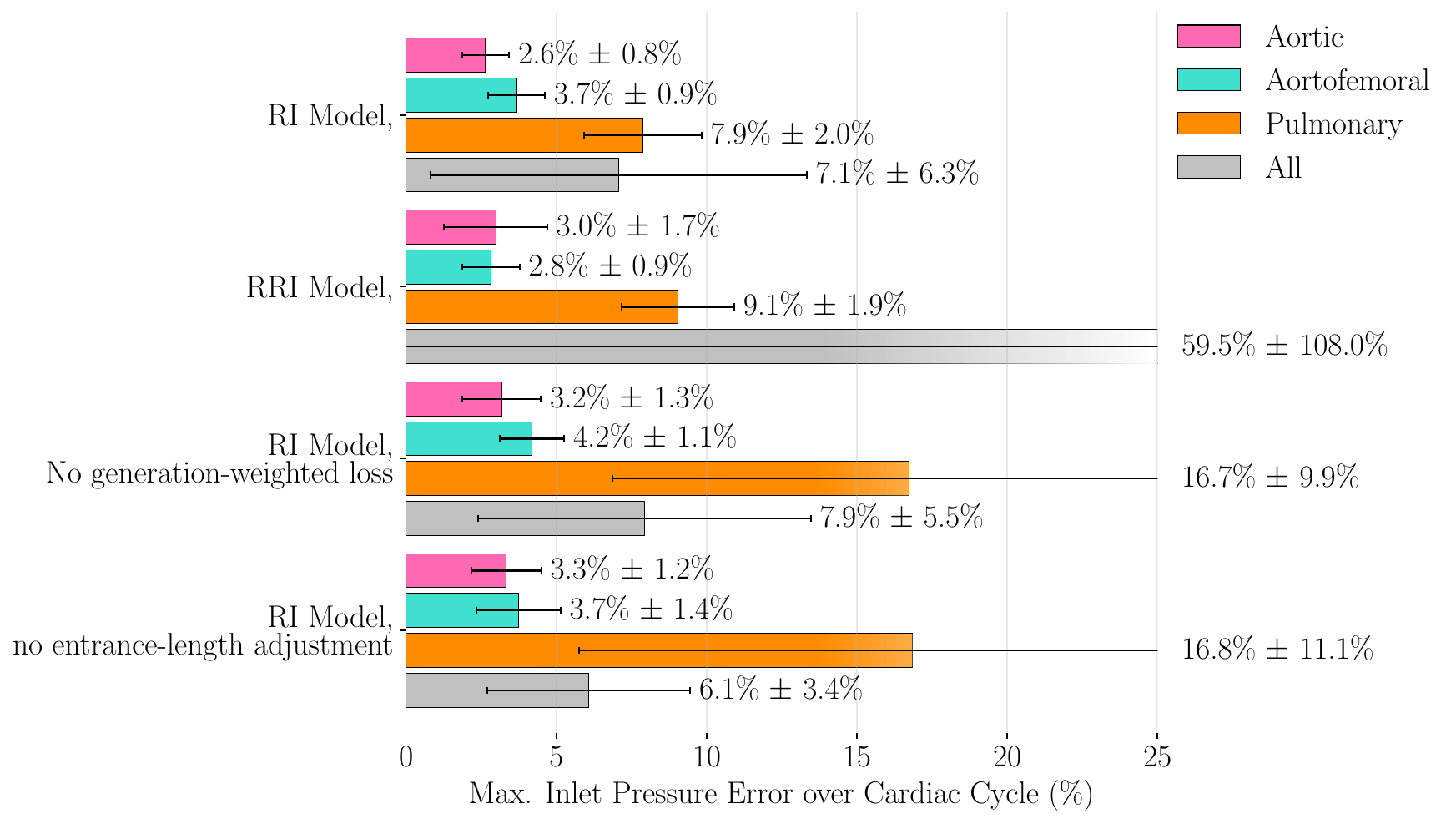}
    \caption{Average and $95\%$ confidence intervals for MPE using both learned junction and vessel parameters, computed across five cross-validation trials, for each anatomical cohort.  Each row of bars shows performance differences associated with different combinations of design choices: inclusion of quadratic resistance (RRI vs RI model), standard vs. generation-weighted neural network optimization loss, and entrance length adjustment.}
    \label{fig:error_by_config}
\end{figure}

To test the impact of the proposed design choices on the accuracy of the neural-net predicted 0D models, we run the cross-validation for several different configurations, \Cref{fig:error_by_config}.  Using the RRI model rather than the RI model increases error in the aortic and pulmonary cohorts from $2.6\%$ to $3.0\%$ and $7.9\%$ to $9.1\%$, respectively.  In the aortofemoral cohort, using the RRI model decreases the MPE from $3.7\%$ to $2.8\%$.  Notably, all of these average errors are within a $95\%$ confidence of one another.  For the `all' cohort containing all three anatomical types, the RRI model yields a large average MPE of $59.5\%$, much higher than that of the RI model, $7.1\%$.  All future results discussed in this work use the more robust RI model, unless explicitly stated.

Using the generation-weighted neural network loss improves results across all cohorts.  In the aortic, aortofemoral, and ``all" cohorts, MPE decreases slightly, from $3.2\%$ to $2.6\%$, $4.2\%$ to $3.7\%$, and $7.9\%$ to $7.1\%$, respectively.  The most marked improvement was in the pulmonary cohort, where average MPE is reduced from $16.7\%$ to $7.9\%$.

The entrance-length adjustment has little effect on the aortic, aortofemoral, and ``all" cohorts.  The average MPE for the aortic cohort decreases from $3.3\%$ to $2.6\%$, the average MPE for the aortofemoral cohort remains constant at $3.7\%$, and the average MPE for the ``all" cohort increases from $6.1\%$ to $7.1\%$.  Again, the pulmonary cohort shows the most dramatic improvement, average MPE decreases from $16.8\%$ to $7.9\%$.

\begin{figure}[htb]
    \centering
    \includegraphics[width=0.8\linewidth]{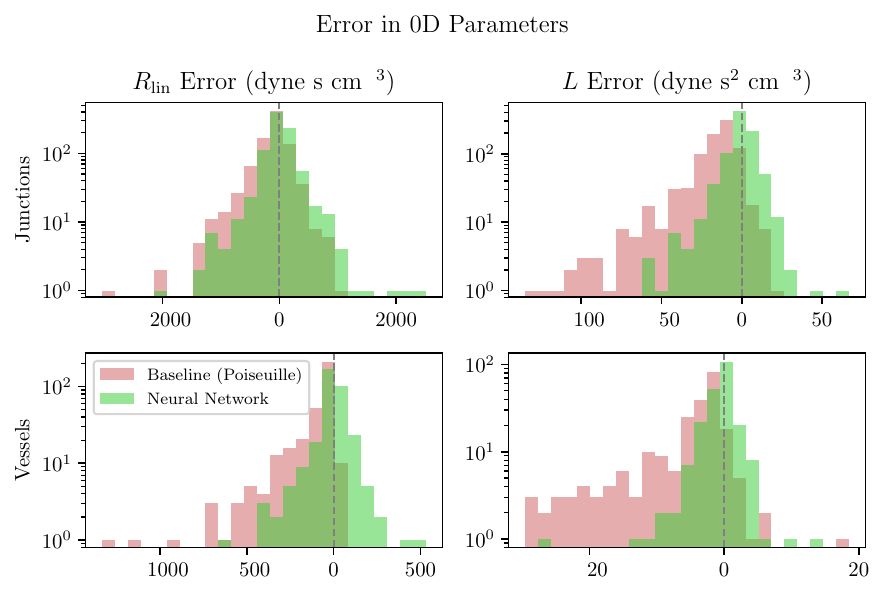}
    \caption{Distributions of errors 0D in parameters predicted by baseline Poiseuille formulas and neural networks.  Error is defined as the difference between the predicted parameter and the optimal parameter found by least squares fit to high-fidelity 3D simulation.  Parameters with values less than five are excluded to avoid misleadingly high relative errors caused by very small optimal parameters.}
    \label{fig:param_comparison}
\end{figure}

We compare the error in the 0D parameters, $R_\text{lin}$ and $L$, predicted using the baseline (Poiseuille assumption) formulas, \Cref{eq:poiseuille_parameters}, and the neural networks in \Cref{fig:param_comparison} and \Cref{tab:param_errors}.  The error is defined as the difference between the optimal (least squares fit) values of the parameter and the predicted parameter.  The baseline formulas tend to under-predict both $R_\text{lin}$ and $L$, while the neural-network errors are more evenly and tightly distributed around zero.  Note that the high relative error for junction $R_\text{Poiseuille}$ in \Cref{tab:param_errors} is a numerical artifact due to a normal-sized error on a relatively small resistance value.

\begin{table}[htbp]
\centering
\begin{tabular}{l r r r r}
\toprule
Parameter & \multicolumn{2}{c}{Relative Error (\%)} & \multicolumn{2}{c}{Absolute Error} \\
 & \underline{Baseline} & \underline{Neural Network} & \underline{Baseline} & \underline{Neural Network} \\
\midrule
Junction $R_{\mathrm{Poiseuille}}$ & 39.20 & 124.72 & $-1.11 \times 10^{2}$ & $2.42 \times 10^{1}$ \\
Junction $L$ & -52.18 & 4.99 & $-1.75 \times 10^{1}$ & $-7.91 \times 10^{-1}$ \\
Vessel $R_{\mathrm{Poiseuille}}$ & -48.64 & 13.11 & $-9.61 \times 10^{1}$ & $-2.66 \times 10^{0}$ \\
Vessel $L$ & -31.11 & -2.37 & $-5.29 \times 10^{0}$ & $-7.11 \times 10^{-1}$ \\
\bottomrule
\end{tabular}
\caption{Average percent error and average error for baseline predicted 0D parameters and neural-net predicted parameters (aggregated across all cohorts). Error is defined as the difference between the predicted and optimal parameter.  Relative errors are divided by the magnitude of the optimal parameter.  Parameters with fitted values with magnitude less than 5 are excluded.}
\label{tab:param_errors}
\end{table}

\section{Discussion}

The cross-validation study, \Cref{fig:cv_max_percent_error_bars}, indicates that neural networks can be effectively used to create 0D models that are significantly more accurate than those that use standard, Poiseuille assumption-based formulas for resistances and inductances.  Even in data-scarce landscapes, the neural network-predicted parameters reduce MPE by $50\%$ or more across all anatomy-specific cohorts.  Analysis of the predicted parameters themselves confirms that neural-network-predicted parameters are more accurate than those predicted by standard Poiseuille assumptions (\Cref{fig:param_comparison}).

Errors are smallest for the simplest anatomical type (aortas) and increase for the more complicated aortofemoral and pulmonary anatomies.  The pulmonary anatomies show particularly high baseline error, a finding consistent with \cite{Pfaller2022AutomatedFlow}, where the authors observe ``pulmonary models exhibit high approximation errors since these models consist of a large number of subsequent vessel junctions, which are currently not modeled in our 1D or 0D models.''  It is not surprising, therefore, that when the junctions are modeled with neural-network predicted parameters, the error decreases about than four-fold.  As pulmonary anatomies are dominated by junctions and feature only short vessel sections (particularly after entrance length adjustment), it is expected that learning vessel parameters has very little effect.  The aortic and aortofemoral geometries, however, contain longer vessel sections and fewer junctions.  For this reason, learning vessel parameters yields a significant increase in accuracy for these geometries.

The cohort containing all geometries shows the most modest improvement in accuracy from using learned junction and vessel parameters, reducing the baseline error by slightly less than half, while the other geometry cohorts all show error reductions of more than half.  This is likely because in the ``all" cohort, the neural networks have a more difficult task in that they must make accurate predictions over a wider range of vessel and junction geometries, without having substantially more data to learn from.  Additional training data and more complex neural networks are likely needed to achieve reductions in error as significant as those observed in the anatomy-specific cohorts.

The configuration comparison \Cref{fig:error_by_config} reveals that the quadratic resistor does not significantly improve results.  Little change is observed in the individual anatomical cohorts, and the performance degrades significantly in the ``all" cohort.  While inclusion of $R_\text{quad}$ slightly improves the 0D model's fit to 3D data \cite{Rubio2025HybridDifferences, Rubio2026Data-drivenModels}, it introduces significant variability and complexity into the task of predicting $R_\text{lin}$ and $R_\text{quad}$.  That is, while the RI model only needs to learn a simple overall relationship between junction geometry and the pressure drop - flow rate dependency, the RRI model needs to tease apart the linear viscous effects in the geometry from the quadratic pressure-recovery or flow separation effects, a more sophisticated and challenging task.  As such, it is more difficult for the neural networks to learn a generalizable relationship between geometry and RRI parameters, compared to RI parameters, especially in a relatively low-data environment. The ``all" cohort contains the most diverse set of anatomies with relatively little training data, so the neural network is expected to perform particularly poorly.  It is also worth noting that the forward 0D solution error is much more sensitive to errors in quadratic resistance than errors in linear resistance, so errors prediction of quadratic resistance degrade the accuracy of the forward simulation much more than similar errors in the prediction of linear resistances.  More complex neural networks trained on larger datasets may prove capable of learning the more complex RRI coefficients with greater accuracy, such that the inclusion of $R_\text{quad}$ indeed yields learned 0D models that consistently produce more accurate 0D results.

The use of the generation-weighted loss, \Cref{eq:prox_loss}, in training the neural network weights is highly beneficial for the pulmonary cohort, while having little effect on the others.  This is expected, as the extensive branching of the pulmonary geometries causes heavy over-representation of distal elements in the training set.  Compared to proximal elements, distal elements have a relatively small effect on overall hemodynamics, so it is less important to predict them accurately.  Without the generation-weighted loss, the neural networks would prioritize accuracy on these common but relatively unimportant distal elements at the expense of accuracy on the more impactful, but less common proximal elements.  By assigning a larger weight to proximal elements, the generation-weighted loss counteracts this deleterious effect.  The other cohorts likely do not benefit as much from the generation-weighted loss because they do not feature as extensive branching and contain more geometries available for training, which may mitigate the issue of proximal element scarcity.

The entrance-length adjustment preprocessing step yields improvements for the aortic and pulmonary cohorts, but had little effect on the aortofemoral cohort.  The benefit of the entrance length adjustment is straightforward.  At the original junction boundaries, which are close to the split site, average pressure $P$ (and therefore $\Delta P$) is quite variable due to local 3D effects, e.g., flow separation, pressure hot-spots at jet-impingement zones, or recirculation zones.  In the context of 0D modeling of bulk parameters, these phenomena amount to noise with little impact on the global system.  (For example, consider a pressure hot-spot at a junction outlet close to the corner that causes a large, negative resistance over the junction.) 

Naturally, it is difficult for a neural network to predict these noisy parameters, and even if it can, they may not actually reflect bulk pressure loss behavior.  In contrast, when the junction outlet points are extended downstream into the outlets, where the flow has developed, we obtain more meaningful measurements of pressure free of local 3D effects. These pressure drops better represent bulk dynamics and have a more direct, learnable relationship to the vascular geometry. On the other hand, extending the junction boundaries to absorb the outlet branches may reduce the resolution of the 0D systems, thereby limiting the ability of the model to represent the 3D flow behavior.  It is possible that in some cases, this effect undoes the benefits of entrance length adjustment for the aortofemoral cohort, or simply that the original junction boundaries are far enough from the split site that extending them further did not make a difference.

\section{Limitations and Future Work}
A primary limitation of this work is the relatively small amount of data used to train the neural networks.  Additional high-fidelity 3D data would provide a more accurate and statistically significant representation of the population from which the neural networks can learn.  Furthermore, additional training data would enable the fitting of more flexible model forms that better capture 3D fluid dynamics.  This additional training data may come from 3D simulations of patient-specific anatomies or synthetically generated anatomies, which are available in larger numbers \cite{Tenderini2025DeformableODEs}.  Additional reduction in error may be achieved by rigorous hyperparameter optimization of the neural networks and fine-tuning of the geometric preprocessing steps, for example, introducing more sophisticated bifurcation-splitting methods and optimizing the choice of pseudo entrance length.

Further interrogation of the data-driven components of this work (the neural networks) should be conducted to increase its interpretability.  Sensitivity analyses or ablation studies may be conducted to better understand the importance of each neural network input feature \cite{Pegolotti2024LearningNetworks, Meyes2019AblationNetworks, Shu2019SensitivityNetworks}.  Deeper analyses may be applied to characterize the role of individual neurons or combinations of neurons in learning physical concepts \cite{MacMillan2025TowardsFeatures, Gurnee2023FindingProbing, Goh2021MultimodalNetworks}.  

Additionally, alternative representations of the system, such as the non-dimensionalization considered in \cite{Rubio2026Data-drivenModels}, may be considered to facilitate the machine learning problem.  To ease the task of learning the quadratic resistance in the RRI model, alternative fitting techniques may be considered where the quadratic term is a correction to the optimal linear (RI) model.  While this approach is not as faithful to the physics of the problem, it may reduce the variability and magnitude of the quadratic resistance, simplifying the machine learning problem and reducing the sensitivity of the forward solution to errors in the neural networks' predictions.

Another limitation is our relatively simplistic representation of the vessel and bifurcation geometries.  We use $\mathcal{G}_\text{J}$ and $\mathcal{G}_\text{V}$ vectors of about 20 geometric features to represent junction and vessel geometries.  This approach is lightweight and interpretable, but the geometry may be represented in more detail by point cloud or shape analysis approaches \cite{Rygiel2023CenterlinePointNet++:Estimation, Sheng2026Geometry-awareHemodynamics, Charles2017PointNet:Segmentation, Sharp2026StatisticalReview}, enabling more accurate prediction of 0D parameters.

The structure of the 0D model also limits accuracy to some degree.  (Note that this work considers only rigid-wall simulations, but may be extended to model fluid-structure interaction by the incorporation of a 0D capacitor to capture wall compliance.) As we see in \Cref{fig:cv_max_percent_error_bars}, even with the optimal parameters, the 0D model cannot exactly replicate the 3D simulation, especially for more complex anatomies.  This may be remedied by finer discretization of the vasculature into 0D elements or inclusion of more complex models for pressure difference in each element, along with improvements to the forward 0D solver to enable it to solve these more complex, nonlinear systems.  As evidenced in the comparison of the RI and RRI models, however, it is more difficult to find and learn the optimal parameters for more complex pressure difference models without additional training data. 

\section{Conclusions}
In this work, we demonstrated the ability of neural networks to predict more accurate parameters for 0D models of patient-specific cardiovascular flows than existing methods based on Poiseuille flow assumptions.  We train neural networks to predict 0D parameters for both vessels and bifurcations by minimizing the difference with respect to optimal 0D parameters determined by fitting the 0D model parameters to the 3D solution.  We run forward 0D simulations for the baseline parameters, neural network-predicted parameters, and optimal parameters, and compare the results to the high-fidelity 3D simulation, finding that the neural network-predicted parameters reduce error by over 50\% compared to the baseline parameters across three types of patient-specific anatomies (aortic, aortofemoral, and pulmonary).

We also introduce and demonstrate the positive effects of several new methodologies supporting our modeling framework.  As preprocessing steps, we present an algorithm to split large multi-outlet junctions into series of bifurcations, an entrance length adjustment to extend the boundaries of the junction region into the outlet branches, where the flow is fully developed.  We propose a generation-based loss for neural network training to remedy the over-representation of distal elements in extensively branched vasculatures.  We also find that a simpler linear model form for pressure loss over vessels and bifurcations generally outperforms the more complex quadratic form given our limited data.

This work improves 0D model accuracy without incurring the computational cost of higher fidelity simulation, bolstering 0D models' utility as real-time digital twins for clinical and engineering applications.  Furthermore, this approach highlights the synergy between machine learning techniques and physics-based approaches in developing accurate, efficient, and interpretable frameworks for characterizing physical systems.

\section{Acknowledgments}
We would like to thank the National Science Foundation Graduate Research Fellowship Program and Grant 2310909, National Institutes of Health Grant R01LM013120, and Stanford Graduate Fellowship for providing funding that enabled this work.  We also thank Dr. Martin Pfaller and his co-authors for sharing the dataset used in \cite{Pfaller2022AutomatedFlow}.

\section{CRediT Statement}
\textbf{Natalia L. Rubio}: conceptualization, data curation, formal analysis, investigation, methodology, software, validation, visualization, writing – original draft. \textbf{ Eric F. Darve}: supervision, writing – review and editing.  \textbf{Alison L. Marsden}: conceptualization, project administration, resources, supervision, writing – review and editing.

\section{Appendices}
\subsection{Neural Network Architecture}
\label{sec:app_nn_architecture}
We report the number of hidden layers and hidden layer width for the neural networks described in \Cref{sec:neural_network}.  The same architectures were used across all anatomical cohorts.

\begin{table}[htbp]
  \centering

  \label{tab:rri-nn-architecture}
  \begin{tabular}{l cc cc}
    \toprule
    & \multicolumn{2}{c}{\textbf{Junctions}} & \multicolumn{2}{c}{\textbf{Vessels}} \\
    \cmidrule(lr){2-3} \cmidrule(lr){4-5}
    \textbf{Model} & Width & \# layers & Width & \# layers \\
    \midrule
    Linear resistor           & 10 & 2 & 10 & 2 \\
    Quadratic resistor & 10 & 2 & 10 & 2 \\
    Inductor                  & 20 & 4 & 10 & 2 \\
    \bottomrule
  \end{tabular}
  \\[0.5ex]
  \caption{Neural network architectures used to predict $R_\text{lin}$, $R_\text{quad}$, and $L$ for vessels and junctions.}
\end{table}

\subsection{Anatomical Cohorts}
\label{sec:app_vmr_names}
Here we provide the names of the patient-specific anatomies used to form the three anatomical cohorts as cataloged in the Vascular Model Repository.  These are drawn from the set of geometries that was used in \cite{Pfaller2021OnSimulations}.  Note that we exclude the pulmonary models from patients with pulmonary hypertension, because, as mentioned in \cite{Pfaller2022AutomatedFlow},  the resistance of the modeled vasculature in these anatomies is negligible compared that of the boundary conditions, such that the hemodynamics are almost entirely dominated by the boundary conditions, yielding artificially small differences between 3D and 0D solutions.
\begin{table}[htbp]
  \centering

  \label{tab:vmr-cohort-geometries}
  \begin{tabular}{@{}lll@{}}
    \toprule
    \textbf{Aortic} & \textbf{Aortofemoral} & \textbf{Pulmonary} \\
    \midrule
    0004\_H\_AO\_SVD & 0028\_H\_ABAO\_H & 0078\_H\_PULM\_H \\
    0005\_H\_AO\_SVD & 0029\_H\_ABAO\_H & 0079\_H\_PULM\_H \\
    0011\_H\_AO\_H & 0030\_H\_ABAO\_H & 0080\_H\_PULM\_H \\
    0012\_H\_AO\_H & 0031\_H\_ABAO\_AAA & 0084\_H\_PULM\_H \\
    0017\_H\_AO\_COA & 0032\_H\_ABAO\_AAA & 0092\_H\_PULM\_H \\
    0021\_H\_AO\_MFS & 0033\_H\_ABAO\_AAA &  \\
    0024\_H\_AO\_H & 0034\_H\_ABAO\_AAA &  \\
    0025\_H\_AO\_MFS & 0035\_H\_ABAO\_AAA &  \\
    0026\_H\_AO\_MFS & 0036\_H\_ABAO\_AAA &  \\
    0027\_H\_AO\_MFS & 0037\_H\_ABAO\_AAA &  \\
     & 0038\_H\_ABAO\_AAA &  \\
     & 0039\_H\_ABAO\_AAA &  \\
     & 0042\_H\_ABAO\_AAA &  \\
     & 0043\_H\_ABAO\_AAA &  \\
     & 0044\_H\_ABAO\_AAA &  \\
     & 0045\_H\_ABAO\_AAA &  \\
     & 0049\_H\_ABAO\_AIOD &  \\
    \bottomrule
  \end{tabular}
    \caption{List of Vascular Model Repository geometries used in each anatomical cohort.}
\end{table}

\subsection{Effect of Empirical Stenosis Correction}
\label{sec:app_empirical_stenosis}
Term 2 of \Cref{eq:vessel_pressure_drop} is an empirical correction for a standard 0D model based on Poiseuille assumptions to account for pressure losses caused by constriction or expansion of a vessel, as in a stenosis.  This appendix shows the difference in 0D error with and without the empirical stenosis correction (as implemented in SimVascular's \href{https://simvascular.github.io/documentation/rom_simulation.html}{\texttt{ROM Simulation Tool}}).

\newpage
\bibliographystyle{unsrtnat}
\bibliography{main}
\end{document}